\documentclass[aps,prl,%showpacs,
twocolumn,nofootinbib]{revtex4-1}
\pdfoutput=1
\usepackage{graphicx,epsf,amssymb,amsbsy,amsfonts,amssymb,amsmath,mathtools}
\usepackage[hidelinks,breaklinks]{hyperref}
\usepackage[dvipsnames]{xcolor}
\usepackage{tikz-cd}
\usepackage[normalem]{ulem}
\usepackage{xcolor}
\usepackage{adjustbox}

\newcommand{\bea}{\begin{eqnarray}}
\newcommand{\eea}{\end{eqnarray}}

\begin{document}

\title{\Large{Force-free magnetosphere attractors for\\ near-horizon extreme and near-extreme limits of Kerr black hole}}

\author{Filippo Camilloni$^{a,b}$}
%\email{filippo.camilloni@nbi.ku.dk}

\author{Gianluca Grignani$^{a}$}
%\email{gianluca.grignani@pg.infn.it}

\author{Troels Harmark$^{b}$}
%\email{harmark@nbi.ku.dk}

\author{Roberto Oliveri$^{c}$}
%\email{roliveri@fzu.cz}

\author{Marta Orselli$^{a,b}$}
%\email{orselli@nbi.dk}

\affiliation{\vspace{0.2cm}$^a$ Dipartimento di Fisica e Geologia, Universit\`a di Perugia, I.N.F.N. Sezione di Perugia, \\ Via Pascoli, I-06123 Perugia, Italy}
\affiliation{\vspace{0.2cm}$^b$ Niels Bohr Institute, Copenhagen University\\
Blegdamsvej 17, DK-2100 Copenhagen \O, Denmark}
\affiliation{\vspace{0.2cm}$^c$ CEICO, Institute of Physics of the Czech Academy of Sciences,\\
Na Slovance 2, 182 21 Praha 8, Czech Republic}

\begin{abstract}
We propose a new approach to find magnetically-dominated force-free magnetospheres around highly spinning black holes, relevant for models of astrophysical jets. Employing the near-horizon extreme Kerr (NHEK) limit of the Kerr black hole, any stationary, axisymmetric and regular force-free magnetosphere reduces to the same attractor solution in the NHEK limit with null electromagnetic field strength. We use this attractor solution as the universal starting point for perturbing away from the NHEK region in the extreme Kerr spacetime. We demonstrate that by going to second order in perturbation theory, it is possible to find magnetically dominated magnetospheres around the extreme Kerr black hole. 
Furthermore, we consider the near-horizon near-extreme Kerr (near-NHEK) limit that provides access to a different regime of highly spinning black holes. Also in this case we find a novel force-free attractor, which can be used as the universal starting point for a perturbative construction of force-free magnetospheres. 
Finally, we discuss the relation between the NHEK and near-NHEK attractors.
\end{abstract}

%-------
\maketitle
%\tableofcontents

%-------

%-------

\section{Introduction}
Spinning black holes with a surrounding magnetosphere and accretion disc are believed to drive astrophysical jets.
Theoretical arguments, as well as numerical simulations, suggest that one can model the magnetosphere to a very good approximation using force-free electrodynamics (FFE) in which the plasma component has a negligible contribution to the energy density. The equations for FFE are Maxwells equations supplemented with the force-free (FF) conditions
\begin{equation}
    \label{FFEeqs}
        D_{[\mu}F_{\nu\rho]}=0,\quad D_{\rho} F^{\rho\nu}= j^{\nu},\quad F_{\mu\nu}j^{\nu} =0
\end{equation}
where $F_{\mu\nu}$ is the electromagnetic field strength, $j^\mu$ is the current and $D_\mu$ is the covariant derivative~\cite{Gralla:2014yja}. Even assuming a stationary and axisymmetric magnetosphere that co-rotates with the black hole, the non-linear nature of the FFE equations in a curved background makes them difficult to solve.
The only known class of exact analytical solutions in the background of a spinning Kerr black hole~\cite{MD2005,Menon_2011,BGJ} has null field strength ($F^2=0$) and instead one needs a magnetically dominated field strength ($F^2 > 0$) for modelling astrophysical jets. 
%-------

Blandford and Znajek initiated a perturbative construction of solutions of the FFE equations \eqref{FFEeqs} for slowly spinning Kerr black holes~\cite{Blandford:1977ds}. However, the most interesting regime to study magnetospheres is actually the one with highly spinning Kerr black holes. In fact, several black holes have been observed to be near-extreme~\cite{McClintock_2006,Gou_2011,Gou_2014} and highly spinning black holes drive the most energetic jets
~\cite{Komissarov_2004,Tchekhovskoy:2011zx,McKinney_2012,Penna_2013}.
%-------

Theoretical results have exploited the near-horizon extreme Kerr (NHEK) \cite{Bardeen:1999px} and near-horizon near-extreme Kerr (near-NHEK) \cite{Amsel_2009} geometries describing, respectively, the near-horizon region of an extreme and a nearly extreme Kerr black hole.
Recent works captured signatures of these geometries in magnetospheres, accretion discs and gravitational binary systems~\cite{Lupsasca:2014hua,Compere:2015pja,Compere:2017zkn,Gralla:2017ufe,Gates:2018hub,Gralla:2016qfw,Compere:2017hsi,Li:2020val}.
%-------

In this letter, we announce a new perturbative method to find  FFE solutions for magnetospheres around an extreme and near-extreme Kerr black hole, which could be of astrophysical relevance. 
This is based on the results of the longer paper \cite{CGHOO} which develop this in detail for an extreme Kerr black hole. In this letter, we announce these results and we take the first steps to implement the method for a near-extreme black hole as well.

For the NHEK limit, this works as follows. Starting with a generic stationary, axisymmetric and regular FF magnetosphere around an extreme Kerr black hole, one finds that in the NHEK limit it always approaches the same attractor solution~\cite{Gralla:2016jfc}.
This attractor mechanism means that one can use the attractor solution as a universal starting point for perturbing away from the NHEK region in order to describe the FF magnetosphere in a larger region of space-time. This is particularly relevant as the NHEK attractor solution has a null electromagnetic field, but it has been shown in \cite{CGHOO} that it is possible to attain a magnetically-dominant field strength by performing a perturbation to second order away from the near-horizon region. This shows that even if the attractor solution is null, one can use it to construct magnetically dominated magnetospheres.
%-------

For the near-NHEK limit, we show that there is an attractor mechanism as well, similar to the one of the NHEK limit.
Indeed, we find that any stationary, axisymmetric and regular FF magnetosphere approaches a novel near-NHEK attractor solution.
This attractor is thus the universal starting point for perturbations away from the near-NHEK region. We consider such perturbations up to first order and discuss how the near-NHEK and NHEK attractors are related by taking advantage of the enhanced symmetries in NHEK and near-NHEK geometries. 
%========

        \section{The NHEK attractor}

%=======

To exhibit the NHEK attractor, our starting point is FFE \eqref{FFEeqs} in the background of a Kerr black hole. The Kerr metric in Boyer-Lindquist (BL) coordinates $(t,\phi,r,\theta)$ is
\begin{align} 
    \label{Kerr}
        ds^2&=-\Big(1-\frac{r_0r}{\Sigma}\Big)dt^2-\frac{2r_0r}{\Sigma}a\sin^2\theta dt d\phi+\frac{\Sigma}{\Delta}dr^2 \nonumber\\
        &\quad +\Sigma d\theta^2 
  +\frac{(r^2+a^2)^2-a^2\Delta\sin^2\theta}{\Sigma}\sin^2\theta d\phi^2,
\end{align}
where $M=r_0/2$ and $J=a M$ are the mass and angular momentum of the black hole. Furthermore, $\Sigma=r^2+a^2\cos^2\theta$, $\Delta=(r-r_+)(r-r_-)$, and $r_\pm=r_0/2\pm\sqrt{r_0^2/4-a^2}$ with $r_+$ the location of the event horizon.
The metric \eqref{Kerr} has two commuting Killing vector fields, $\partial_t$ and $\partial_\phi$, associated with stationarity and axisymmetry,  giving the isometry group $\mathbb{R}\times U(1)$.
%-------

One can show that any stationary and axisymmetric FF field strength around a Kerr black hole can be put in the form \cite{Gralla:2014yja}
\begin{equation}
    \label{FBL}
        F=\frac{\Sigma I(\psi)}{\Delta\sin\theta} dr\wedge d\theta + d\psi\wedge \left(d\phi-\Omega(\psi) dt\right)
\end{equation}
where $\psi(r,\theta)$ is the \emph{magnetic flux} through a circular loop of radius $r\sin\theta$ around the rotation axis, $I(\psi)$ is the \emph{poloidal current} flowing through the loop and $\Omega(\psi)$ is the \emph{angular velocity of the magnetic field lines}. 
That $I$ and $\Omega$ depend only on $\psi$ follows from \eqref{FFEeqs}~\cite{Blandford:1977ds}.
We demand $F$ to be regular at the event horizon $r=r_+$, resulting in the Znajek condition \cite{10.1093/mnras/179.3.457}
\begin{equation}
    \label{znajek}
        \left.        \left( I\Sigma-\sin\theta(r^2+a^2)\left(\Omega-\frac{a}{r_0r_+}\right)\partial_\theta\psi \right) \right|_{r=r_+}=0.
\end{equation}
%
%-------

To derive the NHEK geometry, we zoom in close to the horizon of an extreme ($a=M$)  Kerr black hole by introducing the scaling coordinates $(T,R,\theta,\Phi)$ as 
\begin{equation} 
    \label{scaling}
        T = \lambda \frac{t}{r_0},\quad
        R= \frac{2 r-r_0}{\lambda r_0},\quad
        \Phi=\phi-\frac{t}{r_0}.
\end{equation}
By taking the $\lambda\to 0$ limit while keeping the coordinates $T$, $R$, $\theta$ and $\Phi$ fixed we get the NHEK metric~\cite{Bardeen:1999px}
\begin{equation}
    \label{NHEK}
        ds^2=\frac{r_0^2}{2}\Gamma(\theta)\bigg[-R^2 dT^2+\frac{dR^2}{R^2}+d\theta^2+\Lambda^2(\theta)\big(d\Phi+RdT\big)^2\bigg]
\end{equation}
where $\Gamma(\theta)=(1+\cos^2\theta)/2$ and  $\Lambda(\theta)=\sin\theta/\Gamma(\theta)$.
In the NHEK spacetime the event horizon is located at $R=0$ and the isometry group is enhanced  to $SO(2,1) \times U(1)$. The generators of these isometries in the coordinates \eqref{scaling} read as
\begin{subequations}
    \label{NHEKgenerators}
    \begin{align}
        H_0&=T\partial_T-R\partial_R,\quad Q=\partial_\Phi,\\
        H_+&=\partial_T, \quad H_-=\left(T^2-\frac{1}{R^2}\right)\partial_T-2T R \partial_R-\frac{2}{R}\partial_\Phi.
    \end{align}
\end{subequations}
The algebra is given by $[H_0,H_\pm]=\mp H_\pm$ and $[H_+,H_-]=2H_0$, all the other commutators being trivial.\\
In the extreme case the Znajek condition \eqref{znajek} gives
\begin{equation}
    \label{ZCExt}
        I_0=\frac{\Lambda}{r_0}\left(r_0\Omega_0-1\right)(\partial_\theta\psi)_0
\end{equation}
where $I_0$, $\Omega_0$ and $(\partial_\theta\psi)_0$, refer to $I$, $\Omega$ and $\partial_\theta\psi$ evaluated at the event horizon.
At this stage, $\Omega_0$ and $\psi_0$ are arbitrary functions though we assume $(\partial_\theta\psi)_0\neq 0$
\footnote{The case $(\partial_\theta\psi)_0 =0$ leads to an electrically-dominated solution. Further details are
given in \cite{CGHOO}.}.
%-------

Taking the NHEK limit $\lambda \rightarrow 0$ of the field strength \eqref{FBL} by employing the scaling coordinates \eqref{scaling} and imposing the Znajek condition \eqref{ZCExt}, one finds the following stationary, axisymmetric FF solution in the NHEK geometry
\begin{equation}
    \label{NHEKattractor}
        F_{\text{NHEK}}=\frac{r_0 I_0}{\Lambda} ~d\bigg(T-\frac{1}{R}\bigg) \wedge d\theta. 
\end{equation}
This field strength is null, regular on the future event horizon at $T\to \infty$ and $R\to0$, and self-similar ($F_{\text{NHEK}} \to F_{\text{NHEK}}/c$) under the scalings $T\to T/c$ and $R\to cR$. 
Since one approaches the NHEK field \eqref{NHEKattractor} irrespective of what solution \eqref{FBL} one starts with, we dub this the {\it NHEK attractor solution}.
The solution \eqref{NHEKattractor} was first found in \cite{Lupsasca:2014hua}; later in \cite{Gralla:2016jfc} it was shown using scaling symmetry arguments that the limiting NHEK field must be null and self-similar, and hence equal to \eqref{NHEKattractor}.
%=======

\section{Perturbing away from the attractor}

%======

In this section, we propose our method to construct magnetically-dominated FF magnetospheres. To show how the method works in practice, we focus on the extreme Kerr background as a case study. We anticipate the main results of \cite{CGHOO}, where the step-by-step derivation is provided in great detail.

The attractor solution \eqref{NHEKattractor} is our starting point for a perturbative construction of a FF magnetosphere in the extreme Kerr background, since by including higher powers of $\lambda$ one moves away from the NHEK region.
To this end one can write the $\lambda$ expansion of extreme Kerr metric as $g = \sum_{n=0}^{\infty} \lambda^n g^{(n)}$ where $g^{(0)}_{\mu\nu}$ is the NHEK metric~\eqref{NHEK}.
The expansion of $\psi$ around the event horizon reads as $\psi(r,\theta) = \sum^\infty_{n=0}\frac{1}{n!}\left(r-\frac{r_0}{2}\right)^n (\partial^{(n)}_r \psi)_0= \sum^\infty_{n=0}\frac{1}{n!}\left(\frac{r_0}{2}\lambda R\right)^n (\partial^{(n)}_r \psi)_0$.
Hereafter we adopt the notation that $\psi_n=\psi_n(\theta) := (\partial^{(n)}_r \psi)_0$ is the $n$-th radial derivative of $\psi$ evaluated at the horizon of the extreme Kerr spacetime. Similar expansions hold for $I(\psi)$ and $\Omega(\psi)$.
Using these expansions in Eq.~\eqref{FBL}, one can formally rearrange the field strength $F$ as an expansion in $\lambda$ \cite{Gralla:2016jfc}
\begin{equation}
    \label{Fexpansion}
        F=\sum_{n=-1}^{\infty}\lambda^{n}F^{(n)}.
\end{equation}
We require the FFE equations to be satisfied at each order in $\lambda$.
After imposing the Znajek condition \eqref{ZCExt}, the leading-order term $F^{(-1)}$ is precisely the attractor \eqref{NHEKattractor}.
%-------

The Lorentz invariant $F^2$ has the following expansion $F^2 = \sum_{n=-2}^{\infty} \lambda^{n} (F^2)^{(n)}$. 
While the leading contribution at $n=-2$ is vanishing and the contribution at $n=-1$ will be proven to be zero, the higher order contributions can make $F^2>0$, thus allowing for a magnetically dominated field strength, which is highly relevant for astrophysical applications. 
Indeed, below we find an explicit example of this by pushing perturbation theory to second order.
%-------

The 1st post-NHEK order is given by the field strength $F^{(0)}$, written in terms of the field variables $\psi_1$, $\Omega_1$, and $I_1$.
The FFE equations \eqref{FFEeqs}  lead to $\Omega_1 = (\Omega'_0/\psi'_0)\psi_1$ and $I_1 = (\Lambda/r_0) \{\partial_{\theta} [ ( r_0\Omega_0-1 )\psi_1 ]-\Lambda^2 \Gamma\Omega_0+2/(r_0\Gamma) \}$, with $\psi_1$ obeying
\begin{equation}
    \psi'_1 -\left(\frac{\Lambda'}{\Lambda} + \frac{\psi''_0}{\psi'_0} \right)\psi_1 -\frac{\Lambda \psi'_0}{r_0}\mathcal{G'} = 0,
\end{equation}
where the function $\mathcal{G}$ is defined by
\begin{equation}
    \label{calG}
        \mathcal{G'}=\frac{\Lambda\Gamma}{r_0\Omega_0-1}\bigg(r_0\Omega_0-\frac{2}{\Lambda^2\Gamma^2}\bigg).
\end{equation}
This allows one to write $\psi_1$, $\Omega_1$, and $I_1$ in terms of the functions $\psi_0$ and $\Omega_0$ as
\begin{equation}
    \label{firstPNfunctions}
        \psi_1 = \frac{\mathcal{G}\Lambda}{r_0}\psi'_0,\quad
        \Omega_1 = \frac{\mathcal{G}\Lambda}{r_0}\Omega'_0,\quad
        I_1 = \frac{\mathcal{G}\Lambda}{r_0} I'_0.
\end{equation}
We remark that the Lorentz invariant $F^2$ at this order, $(F^2)^{(-1)}$, vanishes.
%-------

A special case of the 1st post-NHEK perturbation occurs when $\mathcal{G}=0$ which implies $r_0\Omega_0=2/(\Lambda^2\Gamma^2)=2/\sin^2\theta$. This choice corresponds to the field angular velocity of the Menon-Dermer (MD) class of stationary and axisymmetric solutions with null field strength \cite{MD2005,Menon_2011}. The MD class consists of field variables $(\psi_{\text{MD}}(\theta), I_{\text{MD}}(\theta), \Omega_{\text{MD}}(\theta))$ depending on $\theta$ only; in our perturbation scheme, such class of solutions can be obtained by demanding that all the higher-order terms $(\psi_n, \Omega_n, I_n)_{n\geq1}$ vanish. It is easy to show that the MD field strength, given by Eq.~\eqref{FBL} with field variables $(\psi_{\text{MD}}(\theta), I_{\text{MD}}(\theta), \Omega_{\text{MD}}(\theta))$, approaches the NHEK attractor \eqref{NHEKattractor} in the NHEK limit.
%-------

Proceeding analogously at the next order $F^{(1)}$, we can express $\psi_2$, $\Omega_2$ and $I_2$  in terms of $\psi_0$ and $\Omega_0$ by solving the equations of motion.
From~\eqref{FFEeqs} we get
\begin{subequations}
\begin{align}
    \label{Omega2}
        \Omega_2 &= \left[\frac{\psi_2}{\psi'_0} + \frac{\mathcal{G}^2\Lambda^2}{r_0^2}\left(\frac{\Omega''_0}{\Omega'_0} - \frac{\psi''_0}{\psi'_0}\right) \right]\Omega'_0,
        \\
        I_2 &= \left[\frac{\psi_2}{\psi'_0} + \frac{\mathcal{G}^2\Lambda^2}{r_0^2}\left(\frac{I''_0}{I'_0} - \frac{\psi''_0}{\psi'_0}\right) \right]I'_0,
\end{align}
\end{subequations}
and a second-order non-homogeneus linear differential equation for $\psi_2$ 
\begin{equation}
    \label{psi2_eq}
        \psi''_2 + a(\theta)\psi'_2 + b(\theta)\psi_2 + c(\theta) =0. 
\end{equation}
The coefficients in~\eqref{psi2_eq} can be written in terms of the arbitrary NHEK functions $\psi_0$ and $\Omega_0$~\cite{CGHOO} but their expressions are quite involved and therefore we do not report them explicitly here.
The equation above should be studied numerically, but this is beyond the scope of the present letter.
Nonetheless, it is possible to find an analytic exact solution of equation \eqref{psi2_eq} by taking advantage of the arbitrariness of $\psi_0$ and $\Omega_0$. 
As a proof of concept, we introduce the ansatz $\psi_0 = k_0 \int (1-r_0\Omega_0)^{-1} \Lambda^{-3/2}d\theta$, where $k_0$ is a real constant. 
With this relation between $\psi_0$ and $\Omega_0$, one has that $b(\theta)=1/4(a^2(\theta)+2 a'(\theta)+8)$ and the equation \eqref{psi2_eq} becomes a known differential equation, whose most general solution is given by $\psi_2=\left(1-r_0\Omega_0\right)^{-1}\Lambda^{-1/2}\left(\psi^{h}_2+\psi_2^{nh}\right)$; the homogeneous and non-homogeneous parts are, respectively,
\begin{subequations}
\begin{align}
    \label{psi2hom}
        &\psi^{h}_2(\theta)=c_1\cos\sqrt{2}\theta+c_2\sin\sqrt{2}\theta,~~ c_1, c_2 \in \mathbb{R},
        \\
    \label{psi2nonhom}
        &\psi_2^{nh}(\theta)= \cos\sqrt{2}\theta\int c(\theta)~\left(1-r_0\Omega_0\right)\Lambda^{1/2}~\frac{\sin\sqrt{2}\theta}{\sqrt{2}}d\theta 
    \nonumber\\
        &\quad\quad\quad -\sin\sqrt{2}\theta\int c(\theta)~\left(1-r_0\Omega_0\right)\Lambda^{1/2}~\frac{\cos\sqrt{2}\theta}{\sqrt{2}}d\theta.
\end{align}
\end{subequations}
The computational advantages of the ansatz above are not only that of making the equation \eqref{psi2_eq} accessible analytically, but also that of leaving
the field angular velocity $\Omega_0$ arbitrary: one can either choose $\Omega_0$ equal to $\Omega_{\text{MD}} \equiv 2/(r_0\sin^2 \theta)$ and, starting from that, construct radial corrections to the MD class, or one can choose a different function and construct novel perturbative solutions.
As an educated guess \footnote{We remark that with the choice of $\Omega_0$ as in \eqref{OmegaAnsatz}, $\psi$ is regular on the axis up to the second order in $\lambda$. However, both $I_0$ and $\Omega_0$ are singular on the rotation axis, as in the MD solution.
}, we introduce the following class of field angular velocities% parametrized by $\beta \neq 0$
\begin{equation}
    \label{OmegaAnsatz}
        r_0 \Omega_0 = 1 +\frac{\beta}{2}\left( 1- \frac{2}{\Gamma^2\Lambda^2}\right), \quad \beta \in \mathbb R_{\ne 0},
\end{equation}
where $\beta$ serves to explicitly parametrize deviations from $\Omega_{\text{MD}}$.
Usign \eqref{OmegaAnsatz} it turns out that \eqref{calG} becomes
\begin{equation} 
    \label{calGAnsatz}
        \mathcal{G}(\theta)= g - \left(1+\frac{2}{\beta}\right)\cos(\theta),
\end{equation}
where $g$ is an integration constant.
We can then compute the NHEK, 1st and 2nd post-NHEK orders for arbitrary $\beta$ and $g$ \cite{CGHOO}.
%-------

\begin{figure}[ht]
    \centering
    \includegraphics[scale=0.8]{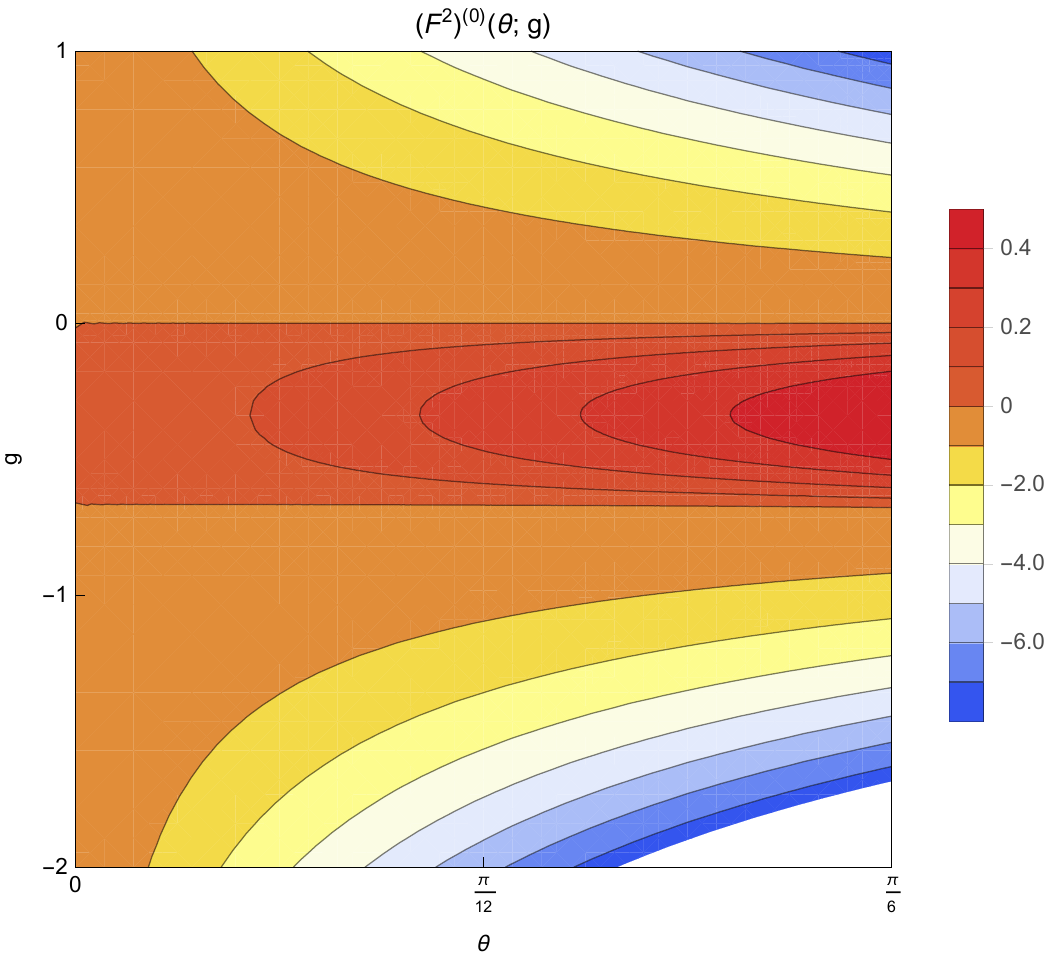}
    \caption{$(F^2)^{(0)}$ with $r_0 =1$, $k_0=2$, $\beta=-2$, $c_1=-g(4-5g)$, and $c_2=0$. The middle strip, defined by $-0.67\lessapprox  g<0$, is the range of values for which $(F^2)^{(0)}$ is positive, \emph{i.e.}, the field strength is a magnetically-dominated solution to FFE.}
    \label{fig:F2contour}
\end{figure}
%-------

In order to show that there exists a class of magnetically-dominated magnetospheres, we now turn our attention to the Lorentz invariant $F^2$. We have already mentioned that the contributions to $F^2$ of the NHEK and 1st post-NHEK orders are vanishing, whereas at the 2nd post-NHEK order it is possible to have $F^2>0$.
The 2nd post-NHEK expression $(F^2)^{(0)}$ is a function of $\theta$ and it depends on the parameters $\beta$, $g$ and the coefficients $c_1$ and $c_2$ in~\eqref{psi2hom}. We fix the coefficients $c_1 = c_1(\beta, g)$ and $c_2=0$ by requiring regularity of $F^2$ at the rotation axis.
For concreteness, let us set $\beta=-2$ equivalent to $\Omega_0=\Omega_{\text{MD}}$. The simplest choice $g=0$ implies that $(F^2)^{(0)}=0$ and the 1st and 2nd post-NHEK field variables vanish, meaning that we reproduce solutions in the MD class up to higher orders. For $g\neq 0$, corresponding to solutions not in the MD class, $(F^2)^{(0)}$ is plotted in Fig.~\ref{fig:F2contour}.
The middle strip, defined by $-0.67\lessapprox  g<0$, is the range where $(F^2)^{(0)}$ is positive and therefore the field strength is magnetically-dominated. 
We have then shown that it is possible, by taking into account post-NHEK corrections to the NHEK attractor, to construct magnetically-dominated solutions in the extreme Kerr background. 
%=======

\section{The near-NHEK attractor}

%=======

We now examine the near-NHEK limit of the general stationary and axisymmetric FF field strength \eqref{FBL} obeying the regularity condition \eqref{znajek}.
The near-NHEK limit of the Kerr metric zooms into the geometry close to the Kerr black hole event horizon $r=r_+$ while at the same time probing nearly extreme spin.
One can perform the near-NHEK limit by introducing 
\begin{equation}
    \label{scalingcoord}
    \begin{aligned}
        &a=\frac{r_0}{2}\sqrt{1-\sigma^2\lambda^2},
    \\
        \tilde{T}=\lambda \frac{ t}{r_0},\quad&
        \tilde{R}=2\frac{r-r_+}{\lambda r_0},\quad
        \tilde{\Phi}=\phi-\frac{t}{r_0},
    \end{aligned}
\end{equation}
Here $\sigma$ parametrizes the deviation from extremality. 
In the scaling limit $\lambda\to0$, performed while holding $\tilde{T}$, $\tilde{R}$ and $\sigma$ fixed, the radial coordinate approaches the horizon, while the spin parameter tends to its extreme value. The result is the \emph{near-NHEK} geometry \cite{Amsel_2009,Bredberg_2010} 
\begin{align}
    \label{nNHEK}
        d\tilde{s}^2&=\frac{r_0^2}{2}\Gamma\bigg[-\tilde{R}(\tilde{R}+2\sigma)d\tilde{T}^2+\frac{d\tilde{R}^2}{\tilde{R}(\tilde{R}+2\sigma)} +d\theta^2 \nonumber\\
        &\qquad\qquad+\Lambda^2\Big(d\tilde{\Phi}+(\tilde{R}+\sigma)d\tilde{T}\Big)^2\bigg].
\end{align}
The event horizon is now located at $\tilde{R}=0$. The near-NHEK and NHEK geometries have the same isometries. In particular, with the near-NHEK coordinates \eqref{scalingcoord}, the generators are given by
\begin{subequations}
\label{nNHEKgenerators}
    \begin{align}
        &\tilde{H}_0=\frac{1}{\sigma}\partial_{\tilde{T}},\quad \tilde{Q}=\partial_{\tilde{\Phi}},
        \\
        &\tilde{H}_{\pm}=\frac{e^{\mp \sigma \tilde{T}}}{\sqrt{\tilde{R}(\tilde{R}+2\sigma)}}\left[\frac{\tilde{R}+\sigma}{\sigma}\partial_{\tilde{T}}\pm \tilde{R}(\tilde{R}+2\sigma)\partial_{\tilde{R}}-\sigma \partial_{\tilde{\Phi}}\right].
    \end{align}
\end{subequations}
The algebra is indeed the same of the NHEK generators, with the only non-trivial commutators given by $[\tilde{H}_0,\tilde{H}_\pm]=\mp \tilde{H}_\pm$ and $[\tilde{H}_+,\tilde{H}_-]=2\tilde{H}_0$.
In the case $\sigma\neq0$ the coordinate choice \eqref{scalingcoord} is not unique and the scaling limit $\lambda\to0$ is not just a coordinate limit since it involves an expansion around extremality. For this reason the near-NHEK geometry is relevant only for near-extreme Kerr black holes.
%------

Expanding the Kerr field strength \eqref{FBL} in powers of $\lambda$ as in~\eqref{Fexpansion} using \eqref{scalingcoord} and imposing regularity of the field at the future event horizon, the leading order contribution $\tilde{F}^{(-1)}$ reads
\begin{equation}
    \label{nNHEK_Attractor}
        \tilde{F}^{(-1)}=\frac{r_0 I_0}{\Lambda}d\left[\tilde{T}-\frac{1}{2\sigma}\log\left(1+\frac{2\sigma}{\tilde{R}}\right)\right]\wedge d\theta.
\end{equation}
This is the {\sl near-NHEK attractor} solution.
It satisfies FFE equations \eqref{FFEeqs}, it is null $(\tilde{F}^{(-1)})^2=0$ and regular at the future event horizon. It is moreover stationary, axisymmetric and self-similar under the scalings $\tilde{T}\to \tilde{T}/c$, $\tilde{R}\to c \tilde{R}$ and $\sigma \to c \sigma$.
We have thus shown that any axisymmetric, stationary and regular magnetosphere around the Kerr black hole asymptotes to the attractor solution \eqref{nNHEK_Attractor} in the near-NHEK limit. The attractor solution \eqref{nNHEK_Attractor} represents a universal feature of such magnetospheres. It is the starting point for any perturbative approach to the magnetosphere of a highly spinning Kerr black hole around the event horizon.
It is important to remark that the near-NHEK attractor \eqref{nNHEK_Attractor} is robust towards the finer details of how one defines the near-NHEK limit through the scaling coordinates \eqref{scalingcoord}.
Moreover, it can be easily checked that the first order corrections in $\lambda$ to the field variables, $(\psi_1,\Omega_1,I_1)$, around the near-NHEK attractor, have exactly the same structure as in~\eqref{firstPNfunctions} for the NHEK case, whereas the second order $\lambda$ corrections are left for future work.
%-------

\begin{figure}[h!]
\adjustbox{scale=0.85,center}{
	\begin{tikzcd}[column sep= scriptsize, row sep=large, cells={nodes={draw=gray}}]
        &F^{(+,n)}=(-1)^n n!\left(T-\frac{1}{R}\right)^{-n}\tilde{F}^{(-1)} \ar[d, shift left, " (H_-)^n "]\ar[r," H_0 "] & (-n) F^{(+,n)}
        \\
         &\ar[u, shift left, "  (H_+)^n "] \tilde{F}^{(-1)}=-\frac{1}{\sigma}\left(T-\frac{1}{R}\right)^{-1}F^{(-1)}\ar[d, shift left, " H_- "] \ar[r," H_0 "] & 0 
        \\
        0&\ar[l, left, " H_+ "]\ar[u, shift left ,"\times" marking, dashed]-\frac{1}{\sigma}F^{(-1)}\ar[d, shift left, " (H_-)^n "]\ar[r," H_0 "] & -\frac{1}{\sigma}F^{(-1)}
        \\
         &\ar[u, shift left, " (H_+)^n "] F^{(-,n+1)}=(n+1)!\left(T-\frac{1}{R}\right)^{n+1}\tilde{F}^{(-1)}\ar[r," H_0 "] & (n+1) F^{(-, n+1)}
    \end{tikzcd}}
    \caption{The chart shows the
    relations generated by the NHEK isometries.
    Starting from the near-NHEK attractor $\tilde{F}^{(-1)}$, written in NHEK coordinates (see also Eq.~\eqref{2Attractors}),
    repeated (Lie derivative) applications of $H_{\pm}$ along the vertical arrows generate descendents $F^{(\pm,n)} \equiv (\mathcal{L}_{H_\pm})^n\tilde{F}^{(-1)}$.
    The horizontal arrows follow the flow of the isometry $H_0$ and allow one to read the conformal weight $h$ of the field $F$ directly from $\mathcal{L}_{H_0}F = hF $.}
    \label{fig:FlowChart}
\end{figure}
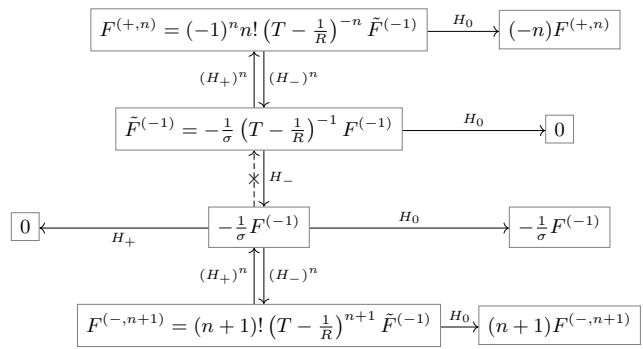

It is interesting to compare the NHEK and near-NHEK attractor solutions. 
There exist three ways to get the NHEK metric~\eqref{NHEK} from the near-NHEK one~\eqref{nNHEK}:
    I. by setting $\sigma=0$ in $\eqref{nNHEK}$, which, according to \eqref{scalingcoord}, consists in \emph{reaching extremality};
    II. by considering $R\gg\sigma$, meaning that near-NHEK is asymptotically NHEK. One can then regard the spacetime of a near-extreme black hole as composed of three different patches: Kerr, NHEK and near-NHEK~\cite{Gralla:2015rpa};
    III. by means of the following \emph{local diffeomorphism} from near-NHEK coordinates $(\tilde{T},\tilde{R},\theta,\tilde{\Phi})$ to NHEK coordinates $(T,R,\theta,\Phi)$~\cite{Compere:2017hsi}
\begin{equation}
\begin{aligned}
    \label{scalingtoscaling}
        &\tilde{T}=-\frac{1}{2\sigma}\log\left(T^2-\frac{1}{R^2}\right),\;\;\,
        \tilde{R}=-\sigma\left(T+\frac{1}{R}\right)R,
    \\
    &\hspace{20mm} \tilde{\Phi}=\Phi+\frac{1}{2}\log\left(\frac{T+1/R}{T -1/R}\right).
\end{aligned}
\end{equation}
We observe that:
    I. by reaching extremality the near-NHEK attractor smoothly reduces to the NHEK attractor,  $\tilde{F}^{(-1)}=F^{( -1)}+\mathcal{O}(\sigma)$;
    II. when $R\gg\sigma$ the asymptotic tensorial structure is the same of the NHEK attractor, $\tilde{F}^{(-1)}\sim (r_0I_0/\Lambda) d(\tilde{T}-\tilde{R}^{-1})\wedge d\theta$;
    III. by using the coordinate transformation \eqref{scalingtoscaling}, one has
    \begin{subequations}
    \label{2Attractors}
    \begin{align}
            \tilde{F}^{(-1)}&=-\frac{1}{\sigma}\left(T-\frac{1}{R}\right)^{-1}F^{(-1)},\\
            \tilde{j}^{(-1)}&=-\frac{1}{\sigma}\left(T-\frac{1}{R}\right)^{-1}j^{(-1)},
    \end{align}
    \end{subequations}
where $j^{(-1)}$ is the vector current associated to the NHEK attractor $F^{(-1)}$.
Therefore the attractor $\tilde{F}^{(-1)}$, which is stationary and axisymmetric in the near-NHEK geometry, also exists as a \emph{non-stationary} field in the NHEK background and viceversa. Eq.~\eqref{2Attractors} also implies that $\tilde{F}^{(-1)}$ and $F^{(-1)}$ can be superposed to generate a new FF solution.\\
The local diffeomorphism \eqref{scalingtoscaling} is also useful to map the generators \eqref{nNHEKgenerators} written in near-NHEK coordinates into the generators \eqref{NHEKgenerators} in NHEK coordinates, according to $\tilde{H}_0\to -H_0$, $\tilde{Q}\to Q$, $\tilde{H}_-\to H_+$ and $\tilde{H}_{+}\to H_{-}$. It is therefore reasonable to outline a full comparison between the two attractors in terms of the isometry $SO(2,1) \times U(1)$. In particular, in the NHEK geometry one derives the relations depicted in Fig.~\ref{fig:FlowChart}.

It is interesting to notice that the isometries $H_\pm$ are able to generate an infinite tower of non-stationary FF fields living in the NHEK geometry. Because of non-linearity, in general, one cannot superimpose FF solutions; nonetheless, analogously to what happens in the case of the two attractors, the superposition of any two of the fields in this tower generates a third non-stationary solution of the system \eqref{FFEeqs} in NHEK. We further observe it is always possible to reach $F^{(-1)}$ by acting with $H_-$ on $\tilde{F}^{(-1)}$, but the converse is never true because the NHEK attractor behaves as a highest weight field, $\mathcal{L}_{H_+}F^{(-1)}=0$. In the near-NHEK geometry a mirrored scheme can be drawn but since from \eqref{scalingtoscaling} one has  $H_{\pm}\to\tilde{H_{\mp}}$, then $F^{(-1)}$ is a lowest weight field.
%======

        \section{Outlook}

%======
This letter announces a new method to construct magnetically-dominated FF magnetospheres around extreme and near-extreme Kerr black holes.
We have shown that there are two null attractor solutions in the NHEK and near-NHEK geometries and that they are the starting point for a perturbative expansion that moves towards a Kerr magnetosphere.
This procedure allows us to obtain magnetically dominated solutions at the second order in the perturbative expansion, as anticipated in this letter and derived in greater detail in \cite{CGHOO}.
%-------

An interesting future direction to consider is the behavior of our magnetosphere solutions near the rotation axis~\cite{GHO2019}.
The analytic solutions we have found, including the solutions in the MD class, have a diverging angular velocity $\Omega$ on the rotation axis. 
Note also that while it can be shown that our perturbative solutions do extract finite angular momentum, the energy extraction near the axis diverges \cite{CGHOO}. The singular behaviour near the rotation axis is due to the specific ansatz we used to solve analytically equation~\eqref{psi2_eq}.
However, in general one encounters light-surfaces before one reaches the rotation axis. Thus, one can possibly have a different solution patch on the other side of the light surface, with matching boundary conditions that do not generate a singular behavior of $\Omega$ on the axis. To analyze this further, one needs to use a matched asymptotic expansion technique~\cite{Armas:2020mio}.
%-------

The novel near-NHEK attractor \eqref{nNHEK_Attractor} describes the universal FF magnetosphere around highly spinning black holes in the near-horizon region. The perturbation scheme introduced in this letter for the NHEK attractor~\eqref{NHEKattractor} can also be used to perturb away from the near-NHEK attractor~\eqref{nNHEK_Attractor}. The advantage of this program is provided by the richer structure of the near-NHEK geometry, which allows for two expansions - one around the horizon and the other away from the extreme spin regime.
This can potentially provide a more accurate magnetically-dominated magnetosphere for highly spinning black holes than with the NHEK procedure.
%-------

Finally, we remark that with our new approach for finding magnetically-dominated magnetospheres for highly spinning black holes, one can presumably calculate the jet power output and study the related energy extraction mechanism from highly spinning black holes, which is relevant for models of astrophysical jets.
%=======

        \section*{Acknowledgments} 

%=======
We thank G.~Comp\`ere, V.~Karas, G.~Menon and T.~Pitik for interesting discussions, and M.~de Cesare to have spotted a misprint. T. H. is supported by the Independent Research Fund Denmark grant number DFF-6108-00340. G. G. and M. O. are supported by the project Fondo Ricerca di Base 2018 of the University of Perugia.  R.O. is funded by the European Structural  and  Investment  Funds  (ESIF)  and  the  Czech  Ministry  of  Education,  Youth  and Sports (MSMT), Project CoGraDS - CZ.02.1.01/0.0/0.0/15003/0000437.

%======

 \bibliography{Bibliography}

%======

\end{document}